# A new method of studying galaxy velocity field: preliminary results

Jacek Chołoniewski ⋆
*Astronomical Observatory of Warsaw University, Aleje Ujazdowskie 4, 00-478 Warsaw, Poland*

**ABSTRACT**
It is shown that the average magnitude – redshift diagram for a magnitude–limited sample of galaxies can be effectively used for mapping galaxy velocity field. This new approach is quantitatively introduced and applied to the CfA, ESO/LV and ZCAT samples of galaxies. Strong arguments are presented for infall velocities in the Virgo and Fornax superclusters and in the Great Wall structure.

## 1 INTRODUCTION

The most popular tool for mapping a galaxy velocity field, and especially its deviation from the universal expansion, is to find distances using the Tully–Fisher relation (for spiral galaxies) and the $D_n - \sigma$ relation (for elliptical galaxies) – see Burstein (1990) and Strauss & Willick (1995) for review.

We propose in this paper a new method of studying the velocity field of galaxies, which takes the observed galactic fluxes (magnitude) and redshifts as an observational database.

This paper presents only the basic idea of the new method. More details about the method and more applications to the observational data will be published in a subsequent paper.

The mathematical formalism used in this paper is closely related to that used by Hendry & Simmons (1990).

## 2 THE BASIC IDEA

If galaxies had the same luminosities ($L = L_o$), their observed fluxes ($f$) would be connected with their distances ($r$) with an unique relation:

$$f = L_o \times \frac{1}{r^2}. \qquad (1)$$

In such a case the flux ($f$) would be an ideal distance indicator.

Unfortunately, this is not true due to two, widely known, reasons:

⋆ E-mail: jch@sirius.astrouw.edu.pl

(i) the luminosities of galaxies have a very wide distribution function, called the luminosity function and denoted by $\Phi(L)$,

(ii) every galaxy sample suffers from observational selection effects (we will consider here the most popular flux–limited case: $f \geq f_{lim}$).

However, if we take into account both of these facts, we can introduce a very similar relation between flux and distance to that which holds for an ideal case (equation 1).

Let us now consider, *the expectation value of the flux for a given distance* $E(f \mid r)$ instead of the flux (as in equation 1). This is the so–called *conditional* expectation value: see eg. Fisz (1963). For a flux–limited sample and under the assumption that the luminosity function is independent of the space coordinates, it is expressed by the formula:

$$E(f \mid r) = \frac{\int_{L_{lim}}^{\infty} L\ \Phi(L)\, dL}{\int_{L_{lim}}^{\infty} \Phi(L)\, dL} \times \frac{1}{r^2} \qquad (2)$$

where:

$$L_{lim} = f_{lim}\ r^2.$$

Equation (2) has a simple intuitive justification: the first factor on the right-hand side is equal to the expectation ("average") value of luminosity at distance $r$; the lower limit of integration in both integrals $L_{lim}$ corresponds to the luminosity cut–off induced by the flux cut-off $f_{lim}$. The integral in the denominator is introduced for normalization purposes.

Equation (2) is valid for an inhomogeneous space distribution of galaxies – see APPENDIX A for the mathematical derivation.

Equation (2) is equivalent to the formulae used by Segal (1989) and Hendry & Simmons (1990) in different contexts.

It is worth noting that equation (2) may by regarded (and used) as a new, inhomegeneity–free tool for the determining of the luminosity function of galaxies.

The magnitude version of the equation (2) has the form:

$$E(m \mid r) = \frac{\int_{\infty}^{M_{lim}} M \, \Phi(M) \, dM}{\int_{\infty}^{M_{lim}} \Phi(M) \, dM} + 5 \log(r) + 25 \qquad (3)$$

where:

$$M_{lim} = m_{lim} - 5 \log(r) - 25$$

and where $m$ and $M$ denote the observed and absolute magnitude respectively.

### 2.1 Special case: Hubble law

In order to confront equations (2) or (3) with observations (namely: magnitude and redshifts for a magnitude–limited sample) we have to derive analogous relations as in equations (2) and (3) but for *redshift* ($cz$).

In order to do this we assume that Hubble Law holds:

$$cz = H_o \, r \quad . \qquad (4)$$

We have, from equation (2):

$$E_H(f \mid cz) = \frac{\int_{L_{lim}}^{\infty} L \Phi(L) \, dL}{\int_{L_{lim}}^{\infty} \Phi(L) \, dL} \times \frac{H_o^2}{(cz)^2} \qquad (5)$$

where:

$$L_{lim} = f_{lim} \, (cz)^2$$

and from equation (3):

$$E_H(m \mid cz) = \frac{\int_{\infty}^{M_{lim}} M \Phi(M) \, dM}{\int_{\infty}^{M_{lim}} \Phi(M) \, dM} + 5 \log(\frac{cz}{H_o}) + 25 \qquad (6)$$

where:

$$M_{lim} = m_{lim} - 5 \log(\frac{cz}{H_o}) - 25.$$

The subscript "H" in equations (5) and (6) indicates that the Hubble Law has been assumed.

### 3 CONFRONTATION WITH OBSERVATIONS

Let us now compare theoretical predictions expressed in equation (6) with the observational data. In order to do this we will compute the average magnitude for a given redshift: $< m \mid cz >$. Since the average value is always an unbiased estimator of the expectation value, both functions: $E_H(m \mid cz)$ and $< m \mid cz >$ should have, within the limits of accuracy, the same shape. Since $E_H(m \mid cz)$ is computed under the assumption that Hubble law holds, any differences between them may be interpreted as a result of non–Hubble motions.

### 3.1 Observations: $< m \mid cz >$

We used as an observational database two magnitude–limited samples of galaxies with redshifts: the CfA sample (Davies & Huchra 1982) for the northern sky, and a subsample of the ESO/LV catalogue (Lauberts & Valentijn 1989) for the southern sky. Both samples are complete to the 14.5 magnitude limit. Magnitudes in CfA are mostly in the Zwicky system ($m_Z$) while for ESO/LV we have choosen the total blue magnitude ($B_T$). Raw data for both samples are presented in Figs 1 and 2.

We have additionally used a subsample of the ZCAT redshift catalogue (Huchra et al. 1993) cut to the 15.5 magnitude limit. Since this subsample is not complete to a specified magnitude limit we will use it in this paper in a limited scope.

As observational representation of the *expectation* value of the observed magnitude $m$ for a given redshift $cz$ – $E(m \mid cz)$ the *averages* of the observed magnitude $< m \mid cz >$ have been computed within 200 km/s bins in $cz$.

The resultant averages are presented in Figs 3 – 7 as one standard deviation error bars connected by broken line.

The same diagrams, namely: the average magnitude – redshift diagrams have been constructed by Segal (1989) but in a different theoretical "environment".

### 3.2 Theory: $E_H(m \mid cz)$

The theoretical shape of the function $E_H(m \mid cz)$ has been computed, using equation (6), assuming the luminosity function in the Schechter (1976) form:

$$\Phi(M) = 10^{0.4(M_* - M)(1+\alpha)} exp(-10^{0.4(M_* - M)}) \qquad (7)$$

with parameters taken from Chołoniewski (1982): $\alpha = -1.1$ and $M_* = -19.2$ for the CfA sample and $M_* = -19.6$ for the ESO/LV sample ($H_o = 100$). The difference between these two $M_*$ values reflects the systematic 0.4 magnitude shift between $m_Z$ and $B_T$ (Huchra et al. 1993).

The resultant function $E_H(m \mid cz)$ is insensitive to the assumed value of the Hubble constant $H_o$.

The shape of the function $E_H(m \mid cz)$ is presented in Figs 3 – 6 as a dotted line. We can easily see that the average magnitude and redshift (or distance) are *positively* correlated.

We have not computed (nor presented) the theoretical curve for the ZCAT sample since this catalogue is not complete to any fixed magnitude limit.

### 3.3 Comparison and interpretation

Both functions: theoretical – $E_H(m \mid cz)$ and observational – $< m \mid cz >$ are presented in Figs 3 (northern sky) and 4 (southern sky).

The approximate agreement between them can be seen. This means that the Hubble law (and other assumptions) are approximately correct. There are, however, some significant differences.

In order to discover the source of these differences let us now study the same diagrams as above but for two pencil–beam subsamples centered on the two neighbouring superclusters: Virgo (Fig.5) and Fornax (Fig.6). Additionally, we present the Great Wall region (Geller & Huchra 1989) using data from the ZCAT catalogue (Fig.7). For both superclusters and for the Great Wall very large differences between theory and observations can be seen. These differences are correlated with the observed number density of galaxies: they are largest where the observed number density is maximal.

The average magnitude and redshift are, in the region of the highest density of galaxies, *negatively* correlated. Since, as have been shown in subsection 3.2, the average magnitude is *positively* correlated with *distance*, it suggests that galaxies which have redshifts *larger* than the average redshift of the supercluster have distances *smaller* than the average distance of the supercluster. And *vice–versa*: galaxies which have redshifts *smaller* than the average redshift of the supercluster have distances *larger* than the average distance of the supercluster.

This can be interpreted as a result of large, non–Hubble, infall velocities in the Virgo, Fornax and Great Wall regions. The amplitude of these velocities should be larger than the supercluster diameter (expressed in velocity terms) wich suggests that a certain fraction of galaxies inside supercluters have negative relative velocities.

## 4   CONCLUDING REMARKS

We have shown that the average magnitude–redshift diagram can provide information about peculiar motion *inside* superclusters of galaxies. The same method can be also used for determining distances to superclusters as a whole.

The approach introduced explicitly controls the selection effects and is insensitive to the inhomogeneities in space distribution. For methods which are based on the Tully-Fisher relation or the $D_n - \sigma$ relation (see Burstein 1990, and Strauss & Willick 1995) these two factors influence the results in a way which requires, in order to make them unbiased, quite complicated statistical treatment (see: Sandage 1994, Hendry & Simmons 1994, Triay, Lachieze-Rey & Rauzy 1994, Hudson 1994)

The method introduced in this paper can be applied to many existing and future magnitude (or diameter) –limited samples of galaxies with redshifts.

We encourage authors who develop models of the galaxy velocity field to produce also (among various others observational predictions) the relation $E(m \mid cz)$ versus $cz$; which can be, as we show in this paper, sensitive to non-Hubble motions. Due to the same reason, we also encourage authors who publish galaxy redshifts (for a complete magnitude–limited samples) to construct average magnitude–redshift diagrams.

One obvious modification of the method proposed in this paper is to use a *diameter*–limited sample instead of a *magnitude*–limited sample, and to replace galaxy *magnitudes* by galaxy *diameters*. Such an approach would be useful especially in a case where one wishes to study large–scale peculiar bulk motions of the superclusters (as a whole) which have large angular separations across the sky. In such cases the average magnitude would be strongly influenced by extinction in our Galaxy, while average *effective* (or characteristic) diameters would be completely insensitive to the extinction.

We have used in our analysis the shape of the luminosity function taken from the literature. It should be remembered, however, that this shape has been computed under the assumption that the Hubble law holds. Since, as we have shown in this paper, there are significant deviations from this law, the shape of the luminosity function should be recomputed without assumption about linear (Hubble) expansion of galaxies. We will try to solve this problem in a future work.

## ACKNOWLEDGMENTS

I thank Jan Zalewski and Krzysztof Jahn for a critical reading of the original version of this paper. This work was partly supported by grant KBN 475/A/95.

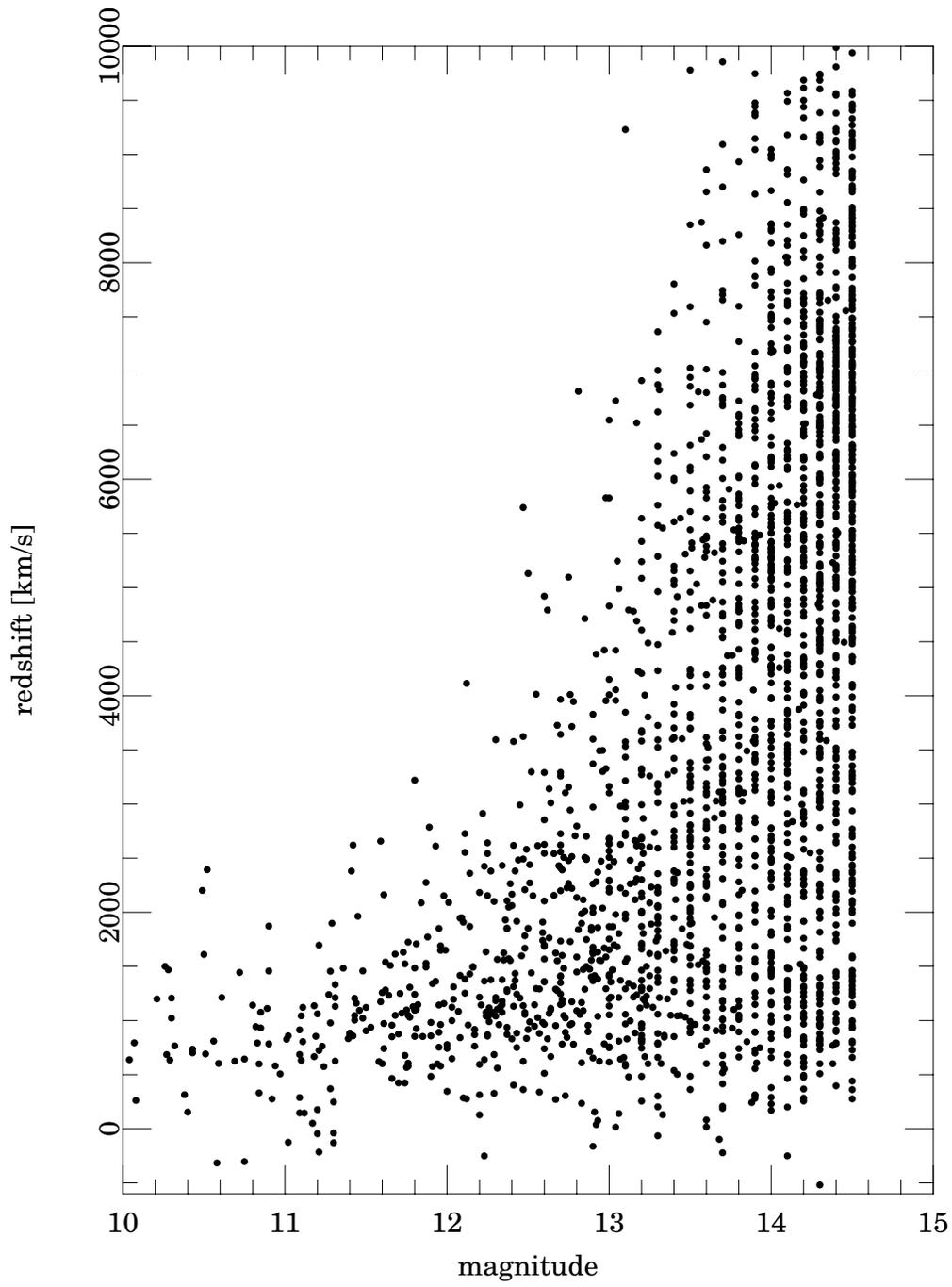

Figure 1. The magnitude–redshift diagram for CfA sample of galaxies.

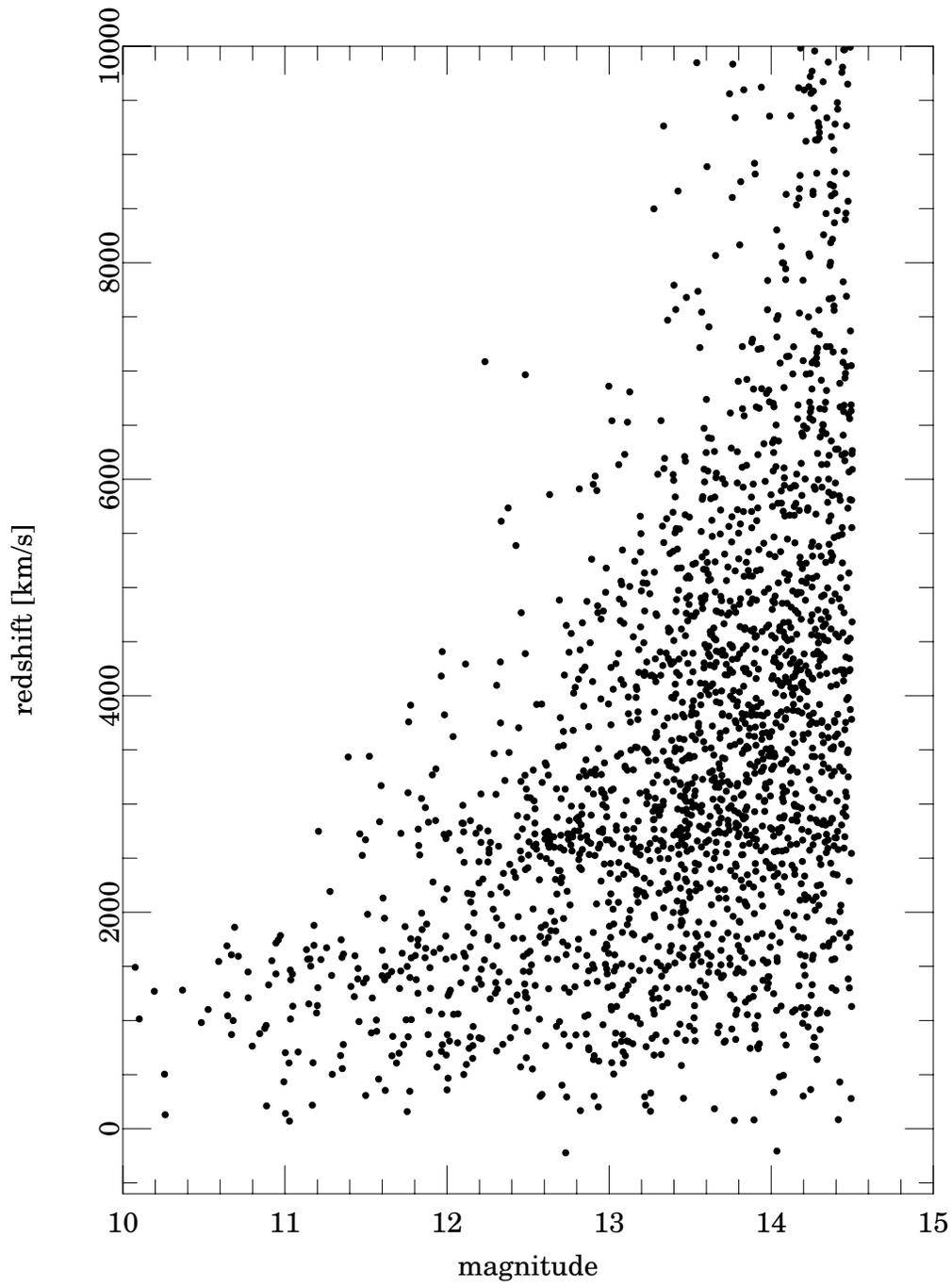

**Figure 2.** The same as Fig. 1 but for ESO/LV sample of galaxies.

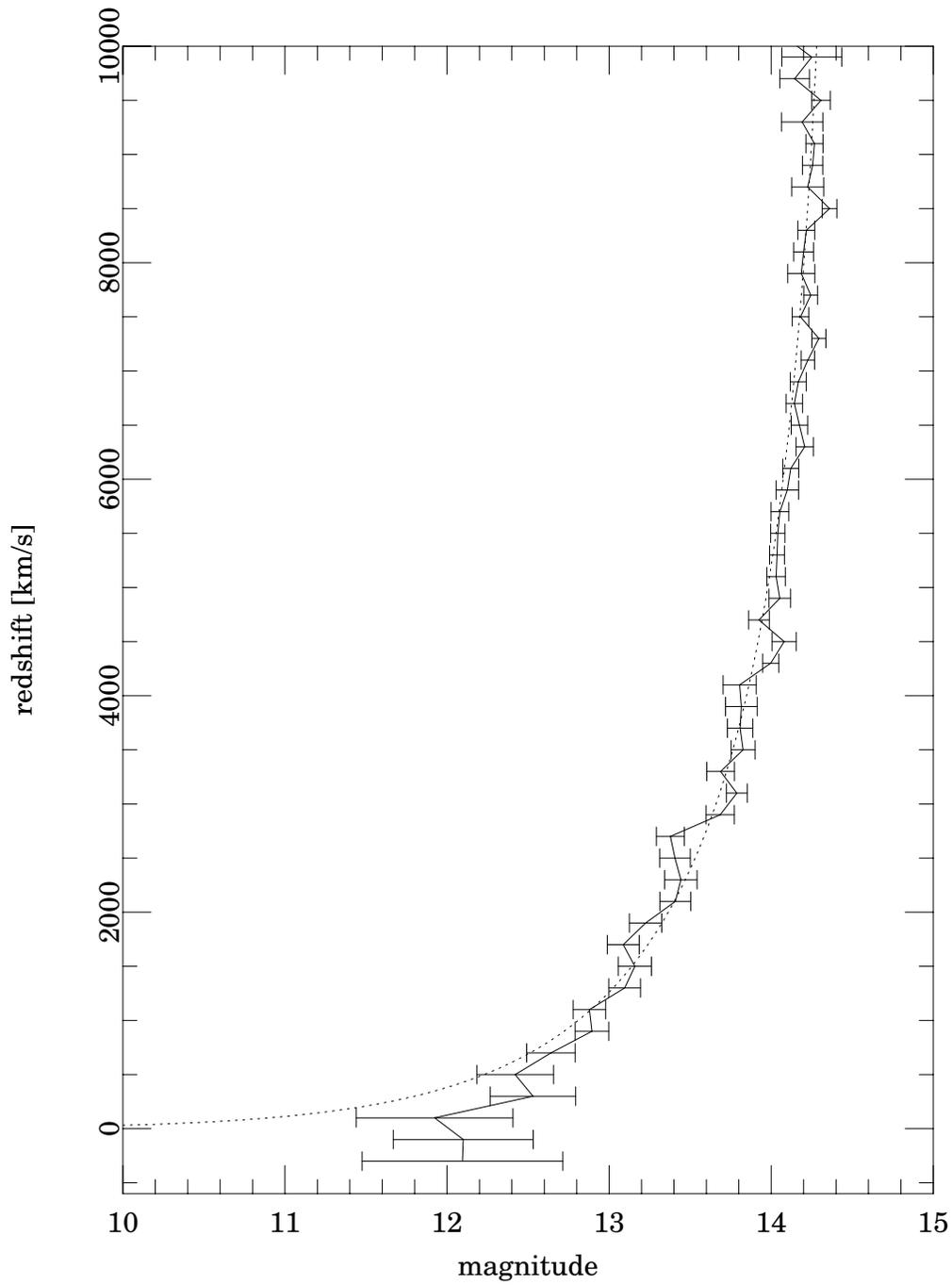

**Figure 3.** The average magnitude as a function of redshift for CfA sample of galaxies. The averages are represented by one standard deviation error bars connected by a broken line. The dotted line represents the expected relation for the pure Hubble law.

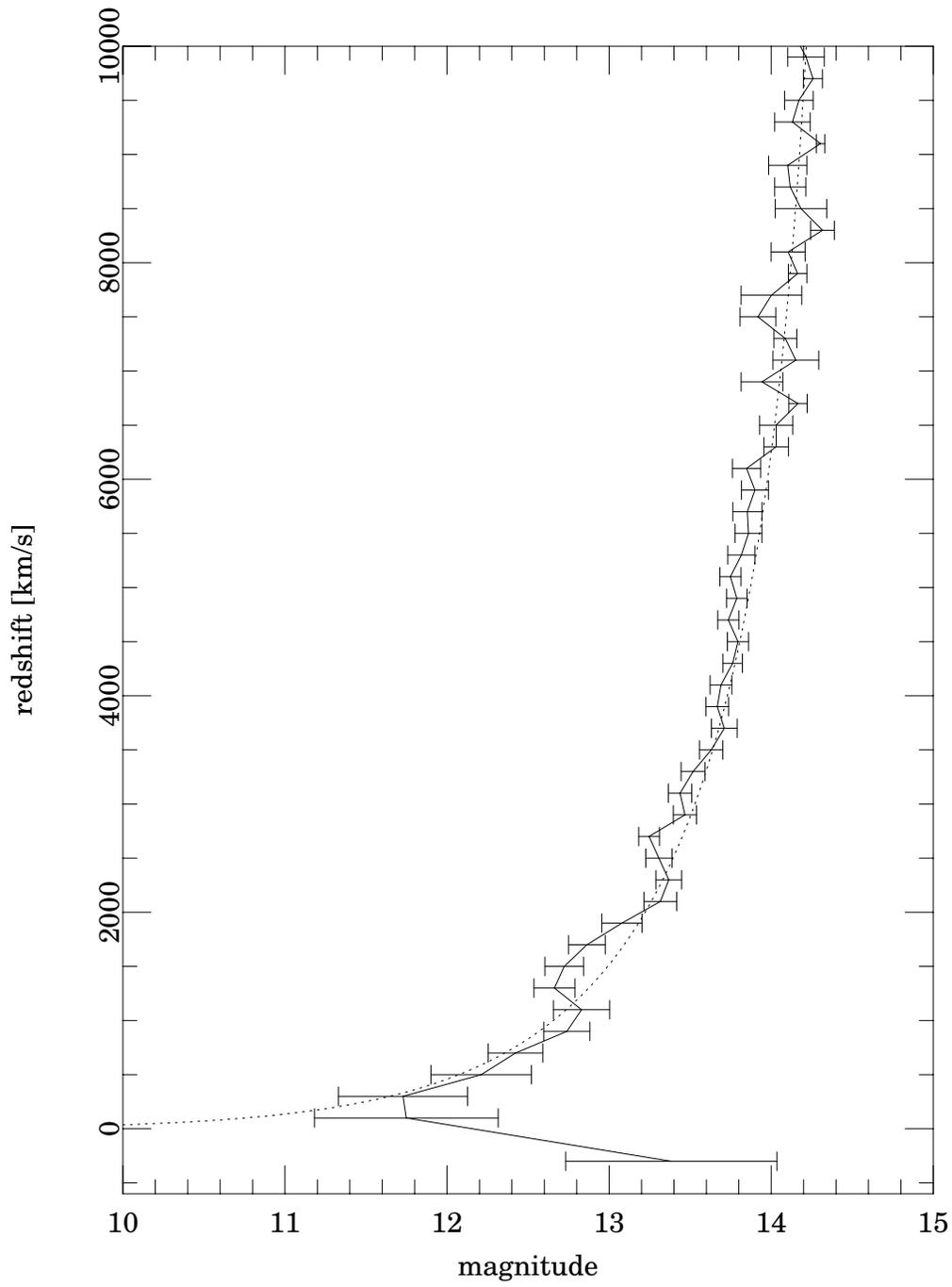

**Figure 4.** The same as Fig. 3 but for ESO/LV sample of galaxies.

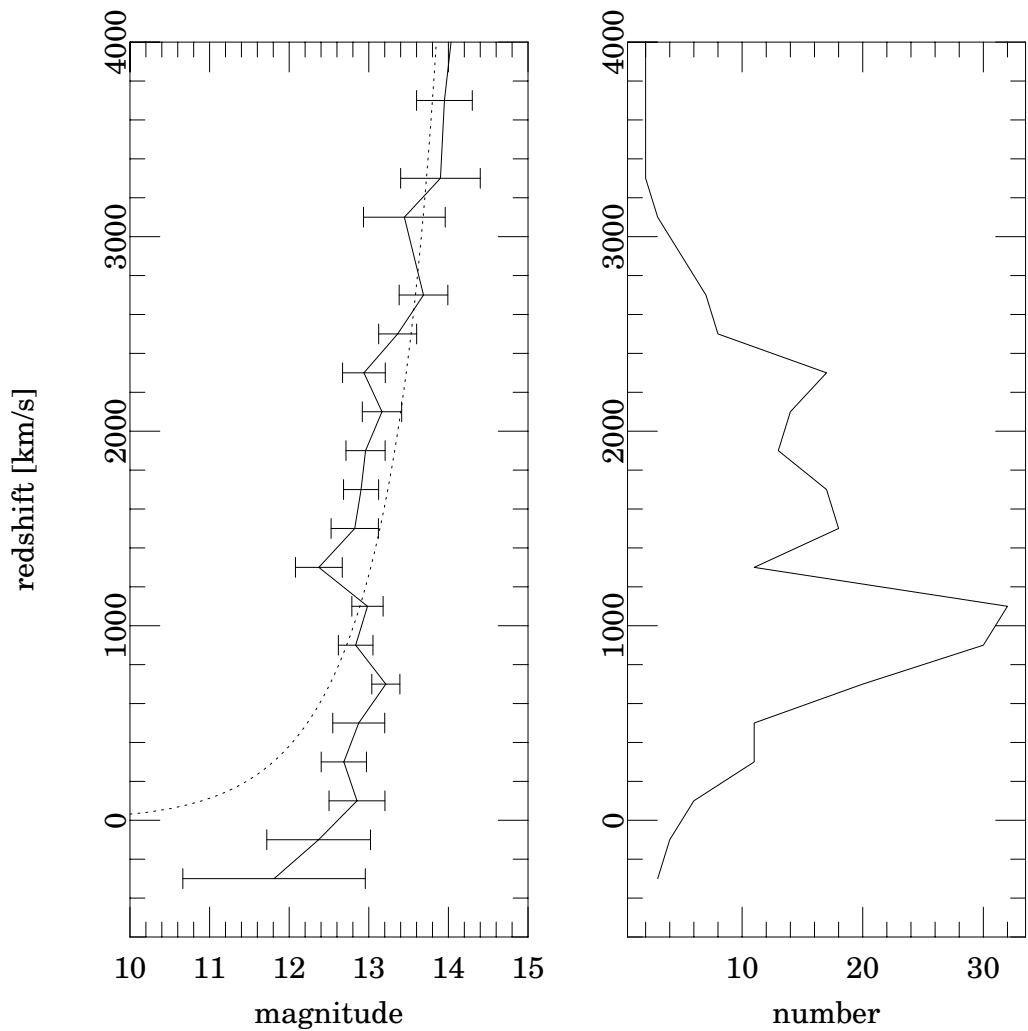

**Figure 5.** Left panel: the same as Fig. 4 but for the Virgo region ($175° < \alpha < 200°$, $0° < \delta < 15°$). The right panel shows the number of galaxies in each, 200 km/s bin used in the left panel. Note that for the Virgo cluster center ($cz \approx 1000\,km/s$) the average magnitude and redshift are negatively correlated. This effect can be explained by infall on to this cluster. The data taken from CfA.

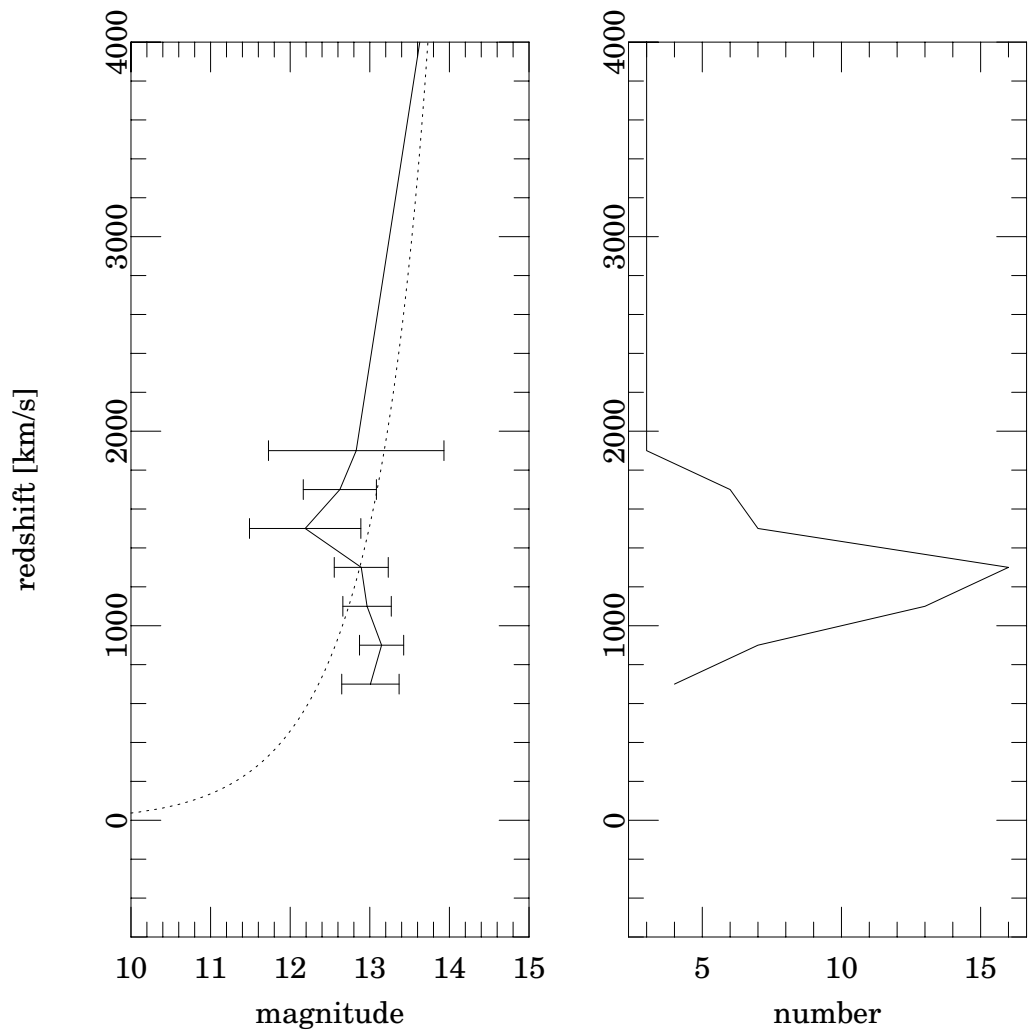

**Figure 6.** The same as Fig. 5 but for the Fornax region ($45° < \alpha < 60°$, $-40° < \delta < -30°$). Note that for the Fornax cluster center ($cz \approx 1300 \, km/s$) the average magnitude and redshift are negatively correlated. The data taken from ESO/LV.

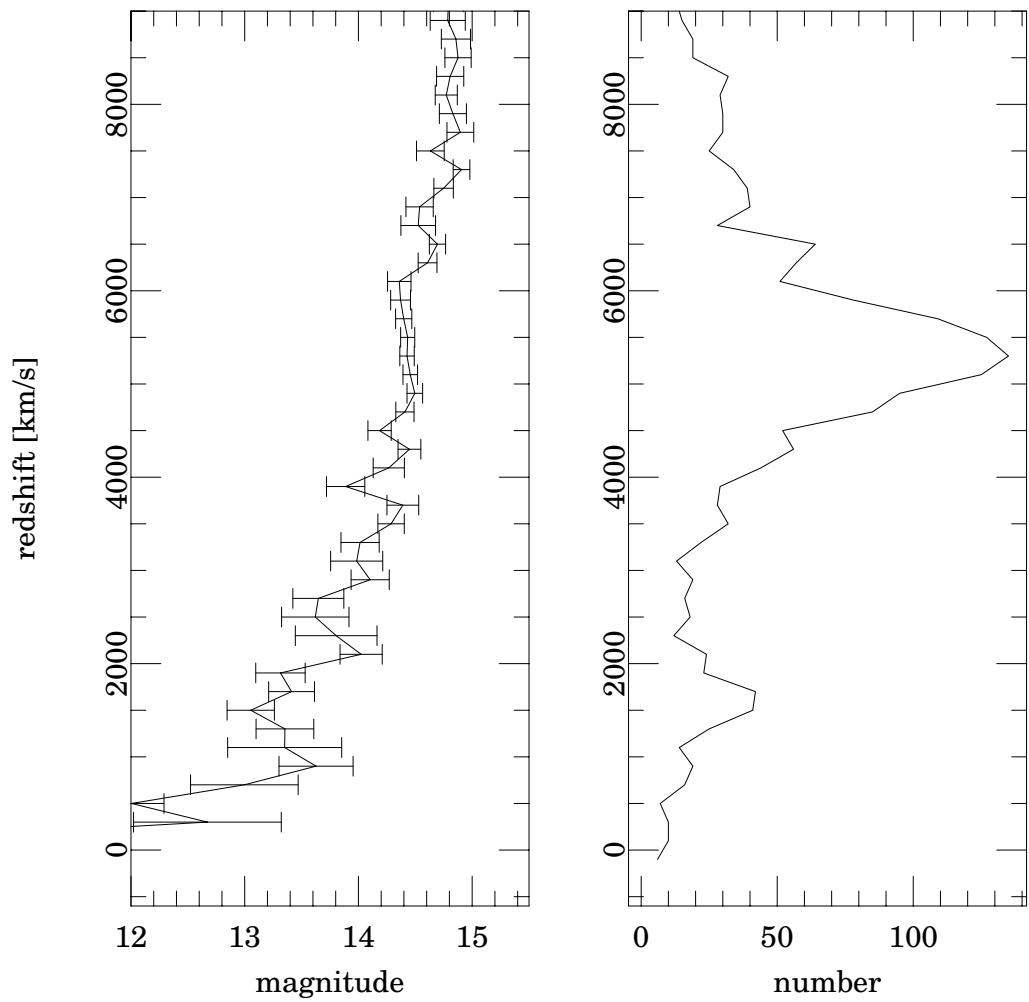

**Figure 7.** The same as Fig. 6 but for the Great Wall region ($0° < \alpha < 50°$, $-90° < \delta < 90°$). Note that for the Great Wall center ($cz \approx 5300\,km/s$) the average magnitude and redshift are negatively correlated. The data taken from ZCAT.

## APPENDIX A: THE DERIVATION OF EQUATION (2)

Let us consider an observed joint distribution function of luminosity ($L$) and space coordinates ($\mathbf{r}$): $N(L, \mathbf{r})$ for a flux–limited sample of galaxies. According to the general rule described by Chołoniewski (1991) this formula can be expressed as:

$$N(L, \mathbf{r}) = \Phi(L, \mathbf{r})\, S(L, \mathbf{r}) \tag{A1}$$

where $\Phi(L, \mathbf{r})$ describes a selection–free distribution function of $L$ and $\mathbf{r}$ and where $S(L, \mathbf{r})$ denotes a selection function – the function which is equal to the probability that a galaxy having properties $L$ and $\mathbf{r}$ belongs to the sample. For flux-limited sample we have:

$$S(L, \mathbf{r}) = I(f \geq f_{lim}) \tag{A2}$$

where the indicator function $I(.)$ is defined as:

$$I(statement) = \begin{cases} 1 & \text{if the } statement \text{ is true} \\ 0 & \text{if the } statement \text{ is false} \end{cases} \tag{A3}$$

Let us assume now that:

(i) $L$ and $\mathbf{r}$ are statistically independent:

$$\Phi(L, \mathbf{r}) = \Phi(L)\, \rho(\mathbf{r}) \tag{A4}$$

where $\Phi(L)$ denotes the luminosity function and $\rho(\mathbf{r})$ denotes the number density of galaxies;

(ii) the space has Euclidean properties:

$$f = \frac{L}{r^2}. \tag{A5}$$

Using these two assumptions in equation (A1) we have:

$$N(L, \mathbf{r}) = \Phi(L)\, \rho(\mathbf{r})\, I(\frac{L}{r^2} \geq f_{lim}). \tag{A6}$$

The next two steps are: the transformation of the variables from rectangular coordinates to spherical coordinates and the integration over angular coordinates which gives:

$$N(L, r) = \Phi(L)\, \rho(r)\, r^2\, I(\frac{L}{r^2} \geq f_{lim}) \tag{A7}$$

where $\rho(r)$ denotes the average number density of galaxies at distance $r$.

Let us now define the conditional distribution function:

$$N(L \mid r) = \frac{N(L, r)}{\int_0^\infty N(L, r)\, dL} \tag{A8}$$

which gives:

$$N(L \mid r) = \frac{\Phi(L)\, I(\frac{L}{r^2} \geq f_{lim})}{\int_0^\infty \Phi(L)\, I(\frac{L}{r^2} \geq f_{lim})\, dL} \tag{A9}$$

and which allows us to compute the conditional expectation value:

$$E(f \mid r) = \int_0^\infty \frac{L}{r^2}\, N(L \mid r)\, dL \tag{A10}$$

and finally:

$$E(f \mid r) = \frac{\int_{L_{lim}}^\infty L\, \Phi(L)\, dL}{\int_{L_{lim}}^\infty \Phi(L)\, dL} \times \frac{1}{r^2} \tag{A11}$$

where:

$$L_{lim} = f_{lim}\, r^2. \tag{A12}$$